\documentclass[twocolumn,showpacs,amsmath,amssymb,bm,amsfonts,floatfix,prl,aps]{revtex4}
\usepackage{graphicx}
\usepackage{color}

\begin{document}
\title{Spectrum for non-magnetic Mott insulators from power functional within Reduced Density Matrix Functional Theory}
\author{Y. Shinohara$^{1}$}
\author{S. Sharma$^{1,2}$}
\email{sharma@mpi-halle.mpg.de}
\author{S. Shallcross$^{3}$}
\author{J. K. Dewhurst$^{1}$} 
\author{N. N. Lathiotakis$^{1,4}$}
\author{E. K. U. Gross$^{1}$}
\affiliation{1 Max-Planck-Institut f\"ur Mikrostrukturphysik, Weinberg 2, D-06120 Halle, Germany.}
\affiliation{2 Department of Physics, Indian Institute of Technology, Roorkee, 247667, Uttarkhand, India.}
\affiliation{3 Lehrstuhl f\"ur Theoretische Festk\"orperphysik, Staudstr. 7-B2, 91058 Erlangen, Germany.}
\affiliation{4 Theoretical and Physical Chemistry Institute, National Hellenic Research Foundation,
Vass. Constantinou 48, GR-11635 Athens, Greece}

\date{\today}

\begin{abstract}

A fully first principles theory capable of treating strongly correlated solids remains the outstanding challenge of 
modern day materials science. This is exemplified by the transition metal oxides, prototypical Mott insulators, that 
remain insulating even in the absence of long range magnetic order. Capturing the non-magnetic insulating state of 
these materials presents a difficult challenge for any modern electronic structure theory. In this paper we demonstrate 
that reduced density matrix functional theory, in conjunction with the power functional, can successfully treat the 
non-magnetic insulating state of the transition metal oxides NiO and MnO. We show that the electronic spectrum retains 
a gap even in the absence of spin order. We further discuss the detailed way in which RDMFT performs for Mott insulators 
and band insulators, finding that for the latter occupation number minimization alone is required, but for the former
full minimization over both occupation numbers and natural orbitals is essential.

\end{abstract}

\maketitle

%%%%%%%%%%%%%%
% Introduction
%%%%%%%%%%%%%%
\section{Introduction}
One of the most useful derivatives of a ground-state density functional theory (DFT) calculation is
the Kohn-Sham (KS) eigenvalues\cite{Kohn1965}, which lead to a \emph{non-interacting} spectrum.
Even though the KS equations represent an auxiliary non-interacting system  
whose states and eigenvalues may be quite different from the true quasi-particle
system,
empirical evidence shows that in many cases this single particle KS
spectrum is in agreement with the x-ray photo-emission Spectroscopy (XPS) and 
Bremsstrahlung isochromat spectroscopy (BIS) experiments\cite{bis,elp91a,elp91b,sawat84}. 
However, for strongly correlated materials this KS spectrum is found to be in fundamental 
disagreement with experimental reality. In the absence of spin-ordering 
all modern exchange correlation (xc) functionals within DFT fail 
to predict an insulating ground-state for
transition metal mono-oxides (TMOs), the prototypical Mott insulators. On the 
other hand, it is well 
known experimentally that these materials are insulating in nature even at 
elevated temperatures (much above the  N\'eel temperature)\cite{tjer,jauch}.
This indicates that magnetic order is not the driving mechanism for the
gap, but merely a co-occurring phenomenon. In fact not only DFT, but most 
modern many-body techniques such as the $GW$ method also fail to
capture the insulating behavior in TMOs without explicit long range spin
ordering \cite{rodl09,arya95,kobayashi08}.

In this regard, the two many-body techniques that are able to capture the correct physics of 
strong correlations are dynamical mean field theory (DMFT)\cite{ren06,kunes08,oki08} and reduced density matrix functional 
theory (RDMFT)\cite{sharma08}; these two methods predicts TMOs as insulators, even in the absence of 
long range spin-order. This clearly points towards the ability of these techniques to capture physics well beyond 
the reach of most modern day ground-state methods.  

Despite this success the effectiveness of RDMFT as a ground state theory has been seriously hampered due to the absence 
of a technique for the determination of the spectral information. 
Recently, this final hurdle has also been removed and the spectral information thus obtained for TMOs was shown to be 
in good agreement with experiments\cite{sharma13}. However, these spectra were calculated in the presence of anti-ferromagnetic order. 
The question then arises as to how effective RDMFT is in describing the insulating state of Mott insulators in the absence of
long range spin order. In order to answer this question, in the present work, we study the spectral 
properties of non-magnetic NiO and MnO. Here former is insulating due to interplay of Mott localization and charge transfer 
effects while the latter is insulating purely due to strong Mott localization.
A  detailed analysis of RDMFT and KS orbitals is performed which shows that, unlike in the case of band insulators, for Mott 
insulators the nature of two set of orbitals are very different and this difference is indeed crucial for the success of 
RDMFT in describing Mott physics.  

\section{Theory}
%%%%%%%%%%%%%%
% RDMFT theory
%%%%%%%%%%%%%%

Within RDMFT the one-body reduced density 
matrix (1-RDM) is the basic variable \cite{lodwin,gilbert} 
\begin{align}
\gamma({\bf r}, {\bf r'})=N\int\!d{\bf r}_2 \ldots d{\bf r}_N
\Phi^*({\bf r'},{\bf r}_2 \ldots {\bf r}_N)\Phi({\bf r},{\bf r}_2 \ldots {\bf r}_N),
\end{align}
where $\Phi$ denotes the many-body wave function.
Diagonalization of this matrix produces a set of natural orbitals\cite{lodwin}, 
$\phi_{j{\bf k}}$, and occupation numbers, $n_{j{\bf k}}$, leading to the spectral 
representation 
\begin{align}\label{srep}
\gamma({\bf r},{\bf r}')=\sum_{j,{\bf k}} 
n_{j{\bf k}} \phi_{j{\bf k}}({\bf r})\phi_{j{\bf k}}^*({\bf r}'),
\end{align}
where the necessary and sufficient conditions for ensemble  $N$-representability
of $\gamma$ \cite{coleman} require $0\le n_{j{\bf k}} \le 1$ for all $j,{\bf k}$, 
and  $\sum_{j,{\bf k}} n_{j{\bf k}}=N$. Here $j$ represents the band index and 
${\bf k}$ the crystal momentum.

In terms of $\gamma$, the total ground state energy \cite{gilbert} of the 
interacting system is (atomic units are used throughout)
\begin{align} \label{etot} \nonumber
E[\gamma]=&-\frac{1}{2} \int\lim_{{\bf r}\rightarrow{\bf r}'}
\nabla_{\bf r}^2 \gamma({\bf r},{\bf r}')\,d^3r'
+\int\rho({\bf r}) V_{\rm ext}({\bf r})\,d^3r \\
&+\frac{1}{2}  \int 
\frac{\rho({\bf r})\rho({\bf r}')}
{|{\bf r}-{\bf r}'|}\,d^3r\,d^3r'+E_{\rm xc}[\gamma],
\end{align}
where $\rho({\bf r})=\gamma({\bf r},{\bf r})$, $V_{\rm ext}$ is a given
external potential, and $E_{\rm xc}$ we call the xc 
energy functional. In principle, Gilbert's \cite{gilbert} generalization of the
Hohenberg-Kohn theorem to the 1-RDM guarantees the existence of a functional
$E[\gamma]$  whose minimum yields the exact $\gamma$ and the exact ground-state
energy of systems characterized by the external potential $V_{\rm ext}({\bf
r})$. In practice, however, the correlation energy is an unknown functional of
the 1-RDM and must be approximated. Although there are several known approximations
for the xc energy functional\cite{M1984,GPB2005,pernal2010,AC3,pade,pnof1,PINOS1,PNOF5,KP2014,localrdmft,piris_jcp2013,GPGB2009,PMLU2012,CP2012,MGGB2013,localrdmftappl}, 
the most promising for extended systems is the power functional\cite{sharma08,sharma13} where the xc energy reads
\begin{align} \label{exc} 
&E_{\rm xc}[\gamma]=E_{\rm xc}[\{\phi_{i{\bf k}}\},\{n_{i{\bf k}}\}] = -\frac{1}{2}\int \, \int d^3r' d^3r
 \frac{|\gamma^{\alpha}({\bf r},{\bf r}')|^2}{|{\bf r}-{\bf r}'|},
\end{align}
here $\gamma^{\alpha}$ indicates the power used in the operator sense i.e.
\begin{equation}
\gamma^{\alpha}({\bf r},{\bf r}')=\sum_i n^{\alpha}_i \phi_i({\bf r})\phi_i^*({\bf r}'), 
\end{equation}
for $\alpha=1/2$ this is the M\"{u}ller functional\cite{mueller}, which is known to severely 
overestimate electron correlation \cite{csanyi,gritsenko,herbert,nekjel} while for $\alpha=1$ this functional is 
equivalent to the Hartree-Fock method, which includes no correlations.  
If $\alpha$ is chosen to be $1/2 < \alpha < 1$, the power functional interpolates between the uncorrelated Hartree-Fock limit
and the over-correlating M\"{u}ller functional.

%%%%%%%%%%%%%%%%%%%
% Technical details
%%%%%%%%%%%%%%%%%%%
All calculations are performed using the full-potential linearized augmented plane wave code Elk\cite{elk},
with practical details of the calculations following the schemes described in Refs.~\onlinecite{sharma08} and 
\onlinecite{sharma13}.

\section{Results}
%%%%%%%%%%%%%%%%%%%%%%%%%%%%%
% Results and discussion: Spectra functional
%%%%%%%%%%%%%%%%%%%%%%%%%%%%%

\begin{figure}[ht]
\centerline{\includegraphics[width=\columnwidth]{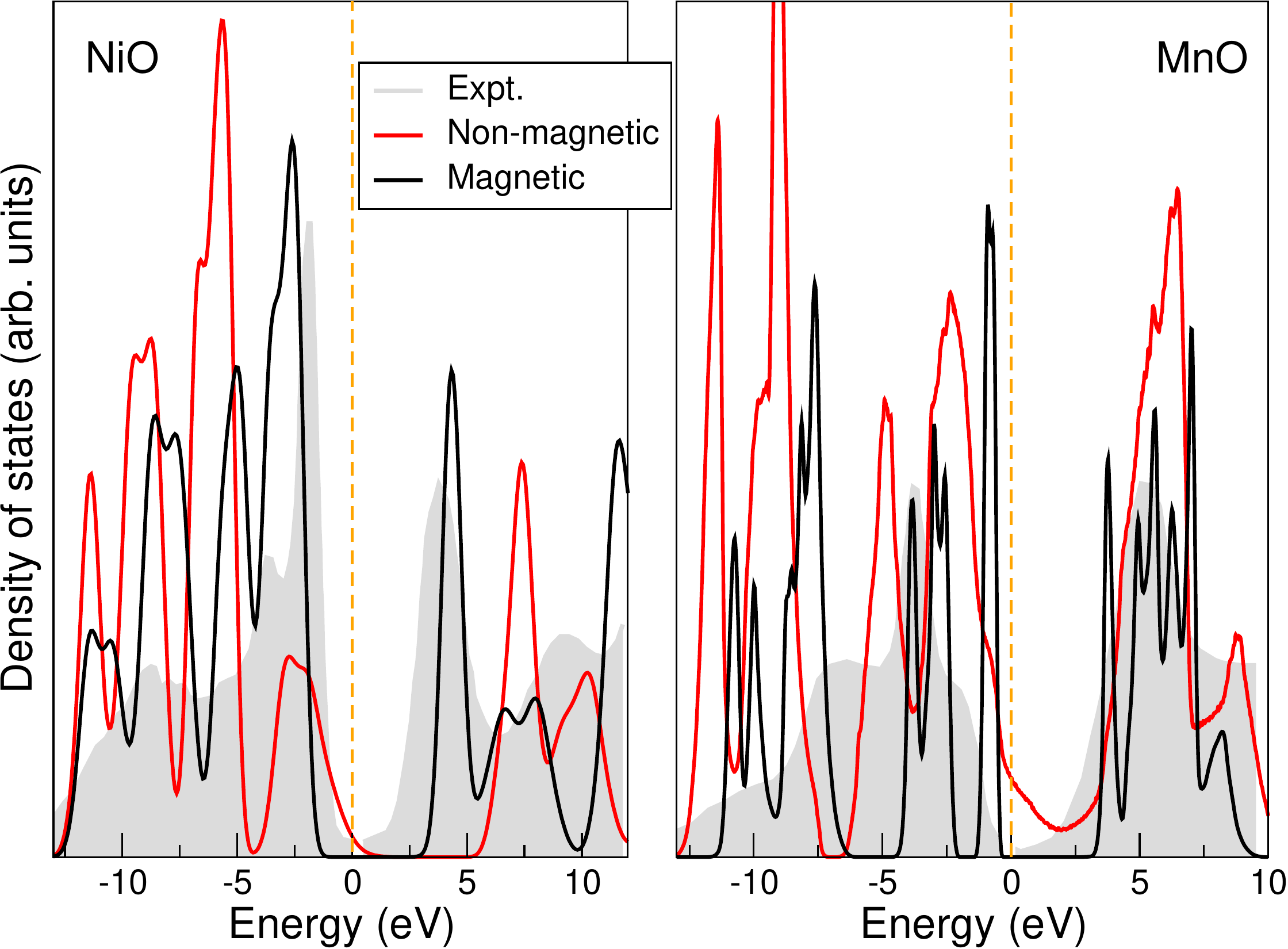}}
\caption{(color online) Density of states as a function of energy (in eV) for NiO (left panel) and MnO (right panel). Results are
obtained with (black) and without (red) long range (anti-ferromagnetic) spin order. For comparison experimental data taken from Refs. \onlinecite{elp91a} 
and \onlinecite{sawat84} is also shown (grey shaded area). Chemical potential is shown at dotted vertical line}
\label{dos}
\end{figure}

Presented in Fig. \ref{dos} are the spectra for the Mott insulators under consideration. It is immediately apparent that RDMFT
captures the essence of Mott-Hubbard physics: both NiO and MnO present substantial gaps at the Fermi energy and are thus insulating in the absence of spin order. 
This fact was already noticed in a previous work\cite{sharma08} in which the presence of gap without any spin-order was 
deduced via very different technique, namely the discontinuity in the chemical potential as a function of the particle number.
A comparison of the non-magnetic spectra with the experimental data shows that the shape of the conduction band is well 
reproduced for both materials, but that the shape of the valence band is not in very good agreement with experiments. This agreement
improves on inclusion of the spin order, indicating that even though insulating nature of TMO's is not \emph{driven} by spin order,
spin polarization significantly effects the spectra of these materials. This is hardly surprising given that NiO and MnO have very 
large local moments of 1.9$\mu_B$ and 4.7$\mu_B$ respectively. 

\begin{figure}
\centerline{\begin{tabular}{c}
\includegraphics[width=0.5\textwidth, clip]{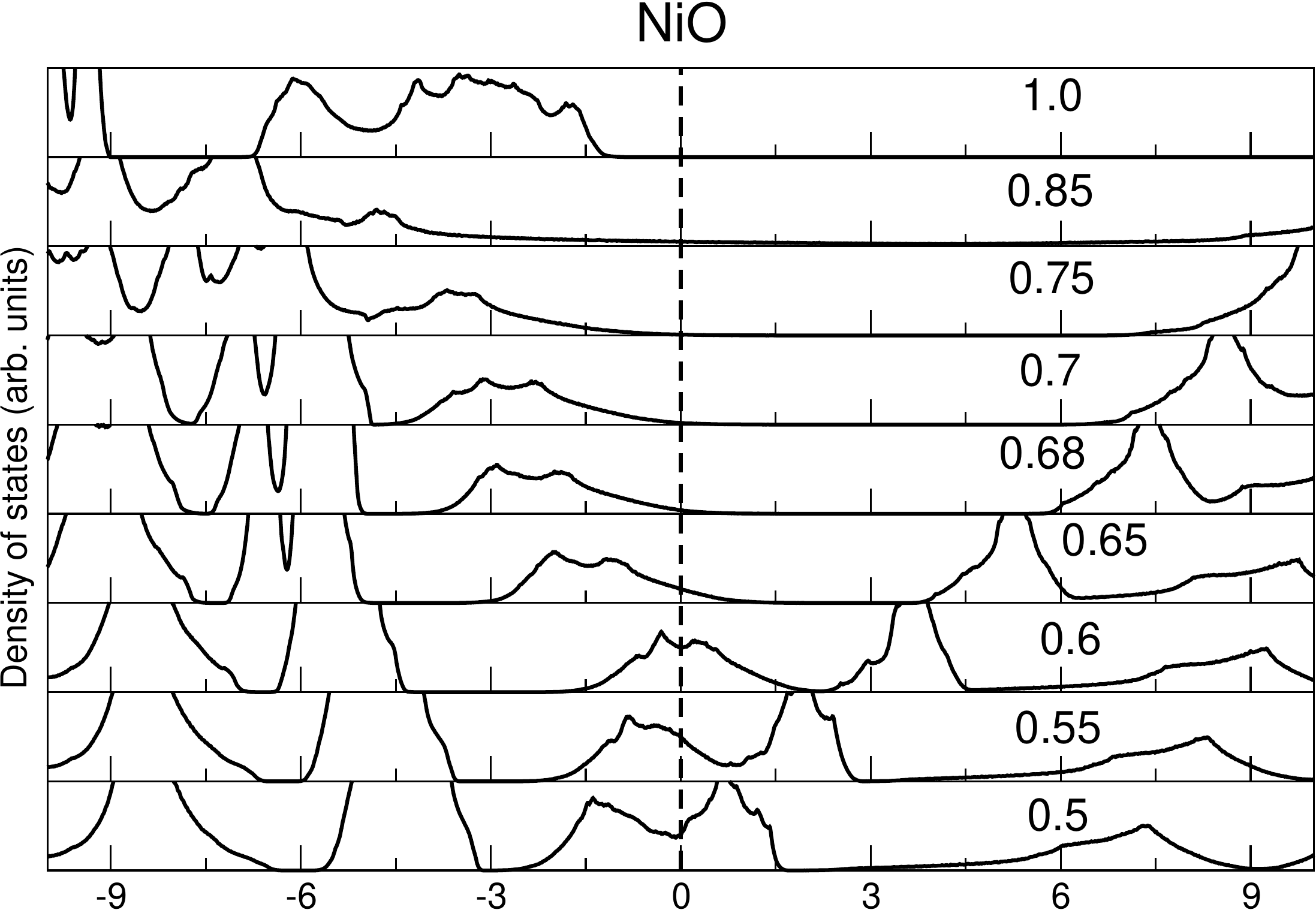}\\
\\
\includegraphics[width=0.5\textwidth, clip]{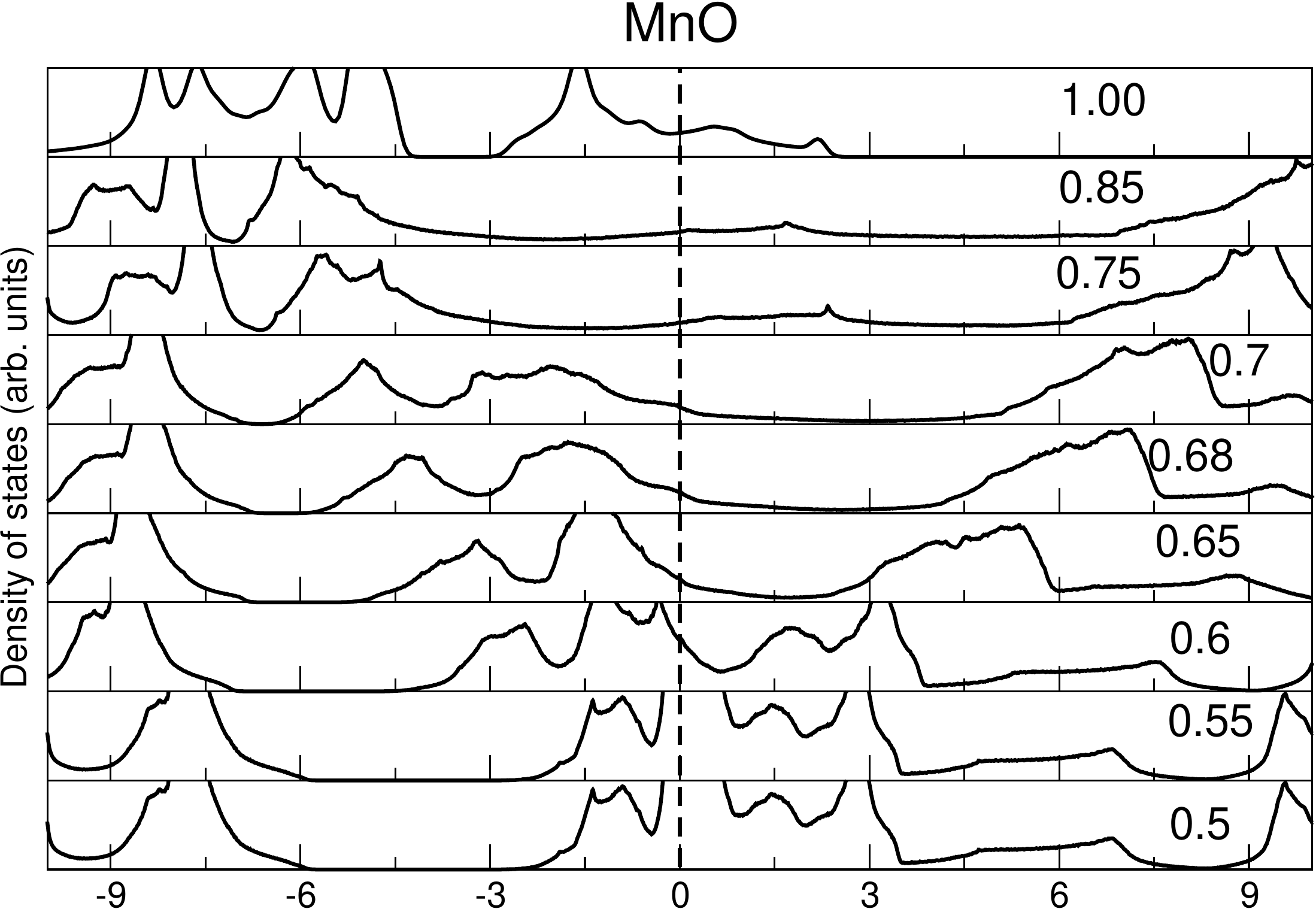} \\
\\
\includegraphics[width=0.5\textwidth, clip]{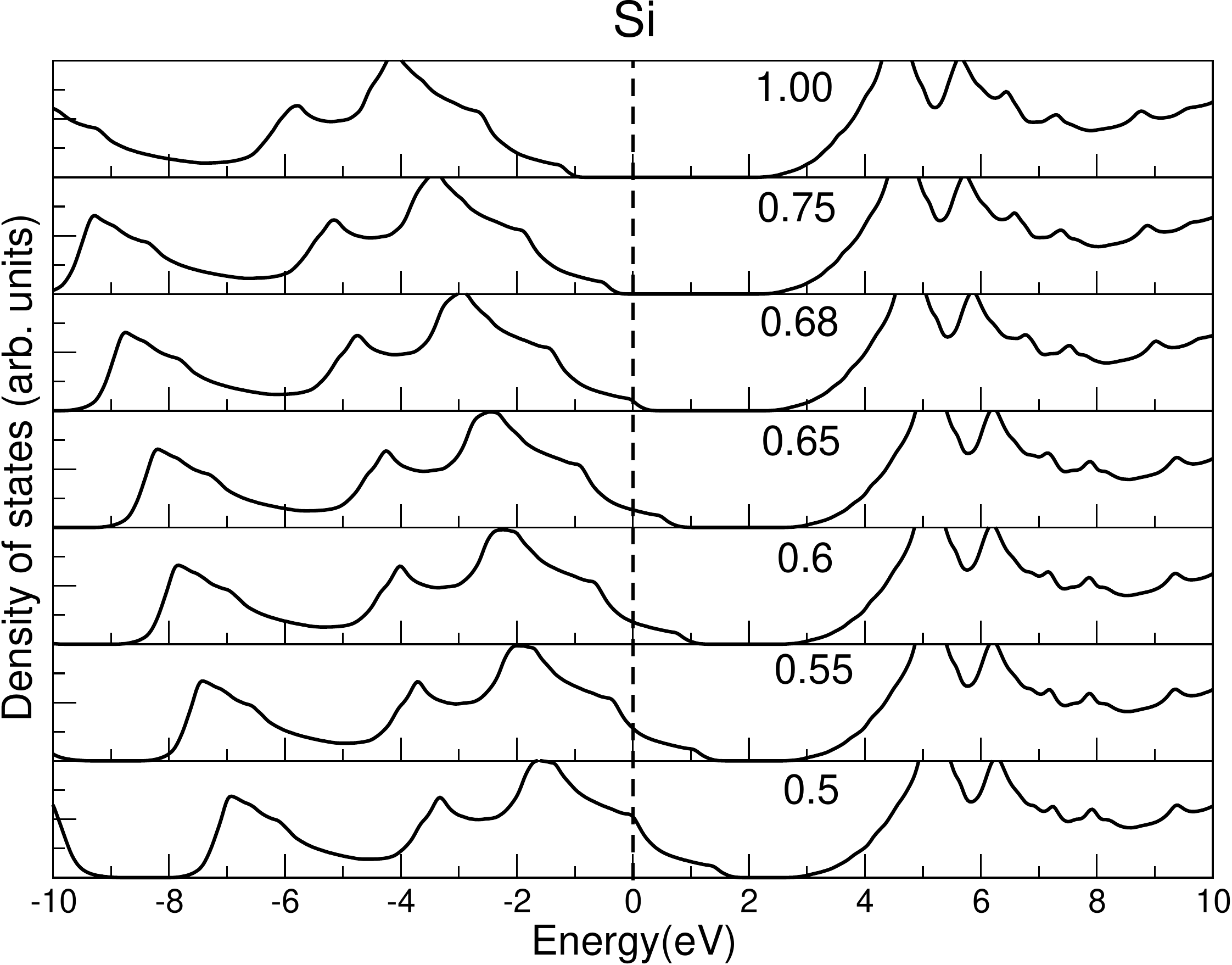} 
\end{tabular}}
\caption{\label{dos-a} Density of states as a function of energy (in eV) for NiO (upper panels), MnO (middle panels),
and Si (lower panels). The results are obtained using different values of $\alpha$ in Eq. \ref{exc}.}
\end{figure}
Correct treatment of correlations is crucial for TMOs, the prototypical strongly correlated materials. As mentioned above 
the power functional interpolates between two limits -- the highly over correlated M\"uller ($\alpha=0.5$) and totally 
uncorrelated Hartree-Fock ($\alpha=1$). We now look at the effect of correlations, by varying $\alpha$, on the spectra of 
Mott insulators (NiO and MnO) and band insulator (Si), see Fig. \ref{dos-a}. The behaviour of the spectra as a function of
$\alpha$ is rather trivial for band insulator, Si; the valence bands rigidly shift lower in energy leading to increase in
the band gap. The behaviour for Mott insulators is  different in that the shape of the bands change as a function of 
$\alpha$. Both for NiO and MnO over correlated M\"uller functional incorrectly gives a metallic ground-state. For NiO, which 
has even number of electrons in a unit-cell, the Hartree-Fock method leads to a very large band gap insulator. In contrast to this,
for MnO, with odd number of electrons in the unit-cell, a single particle theory such as Hartree-Fock can only give rise to a metallic ground state.
This leads to highly non trivial behaviour for MnO as a function of $\alpha$, which must lie within a small range (between 0.65 and 0.7) 
in which the correct insulating ground-state is obtained. Reassuringly, this is also the range of $\alpha$ in which correct ground state behaviour is seen for NiO.

\begin{figure}[ht]
\centerline{\includegraphics[width=0.9\columnwidth]{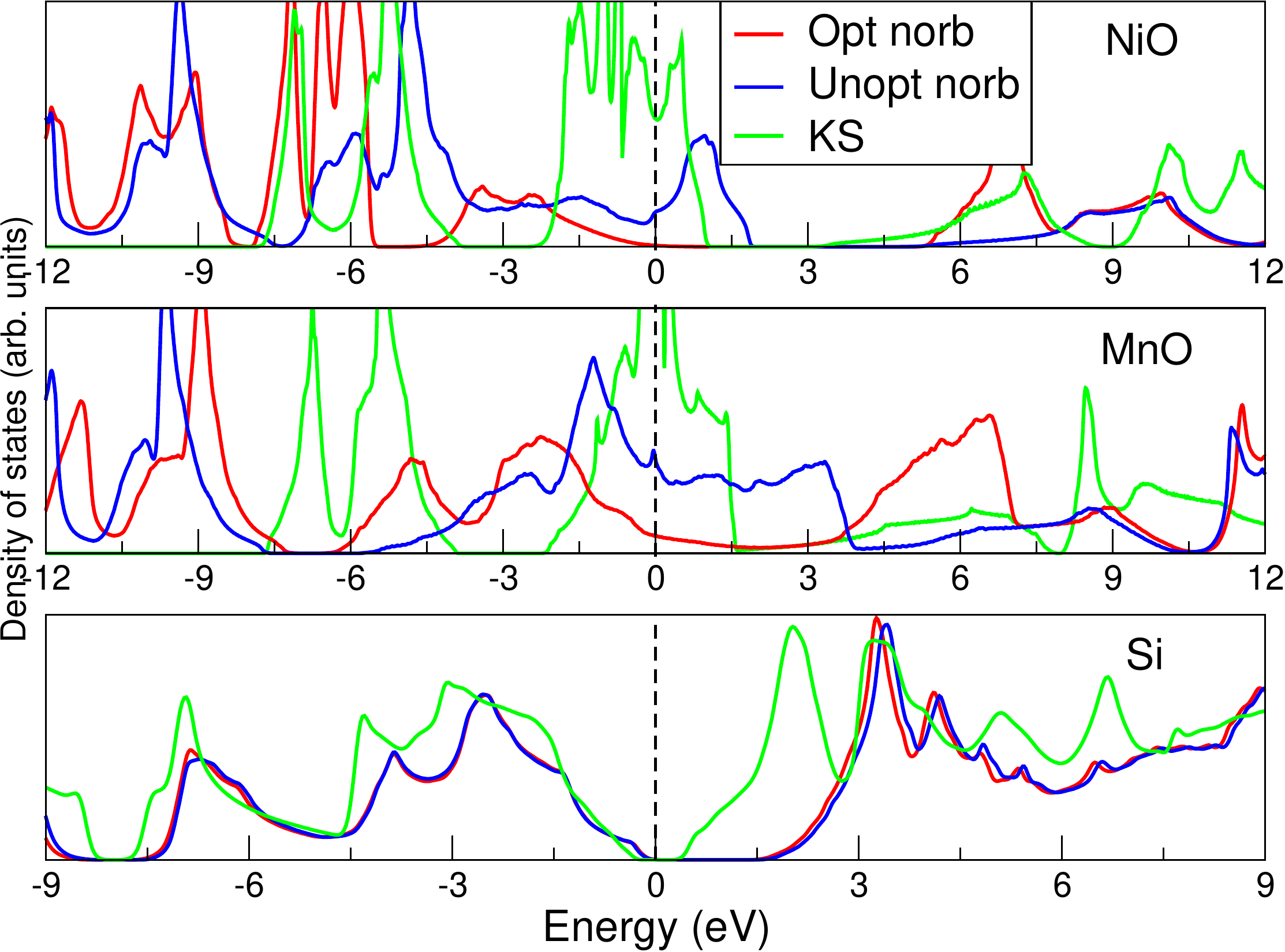}}
\caption{Density of states as a function of energy (in eV) for NiO (top panel), MnO (middle panel) and Si (lower panel). 
Results are obtained with (black) and without (red) optimization of the natural orbitals with in RDMFT. KS results (green) are obtained 
using local density approximation\cite{lda}.}
\label{wfm}
\end{figure}
Within RDMFT there are no Kohn-Sham-like equations to solve, and a direct minimization over
natural orbitals and occupation numbers is required while maintaining the
ensemble $N$-representability conditions. The minimization over occupation numbers is computationally very
efficient (for details see Ref. \onlinecite{sharma08}), but the same cannot be said about the minimization over the natural orbitals.
In practical terms, the natural orbitals (see Eq. (\ref{srep})) are expanded in a set of previously converged 
KS states, and optimization of the natural orbitals is performed by varying the expansion
coefficients. This procedure allows us to examine how different KS states are from fully optimized natural orbitals.
In the present work these KS states were obtained using local density approximation (LDA)\cite{lda}.

In Fig. \ref{wfm} three set of results are shown; (i) KS density of states, (ii) RDMFT density of states obtained without optimizing 
the natural orbitals i.e. by using KS orbitals as natural orbitals but fully optimizing the occupation numbers and  (iii) the fully
optimized RDMFT results i.e. full optimization over the natural orbitals and occupation numbers. From these results it is 
clear that for the band insulator Si it is sufficient to optimize the occupation numbers to increase the band gap in line with experiment;
the KS states are evidently already a very good representation of the natural orbitals. These results are in line with 
our experience with finite systems which shows that orbital optimization results roughly up to 25\% of the 
total correlation energy and the rest 75\% comes from the occupation numbers optimization.
As may be seen in Fig.~\ref{wfm} the opposite situation holds for the 
case of the Mott insulators NiO and MnO: clearly the KS states differ profoundly from the natural orbitals. In this case it is crucial 
to optimize the natural orbitals. The reason for this is that in the case of Mott insulators it is the localization of electrons
which leads to formation of the gap and KS orbitals are not sufficiently localized, thus optimization over the natural orbitals is required.

%%%%%%%%%%%%%%%%%%%%%%%%%%%%%%%%%%%%%%%%%
% Results and discussion: Charge density
%%%%%%%%%%%%%%%%%%%%%%%%%%%%%%%%%%%%%%%%%
\begin{figure}[ht]
\centerline{\includegraphics[width=0.9\columnwidth,angle=-0]{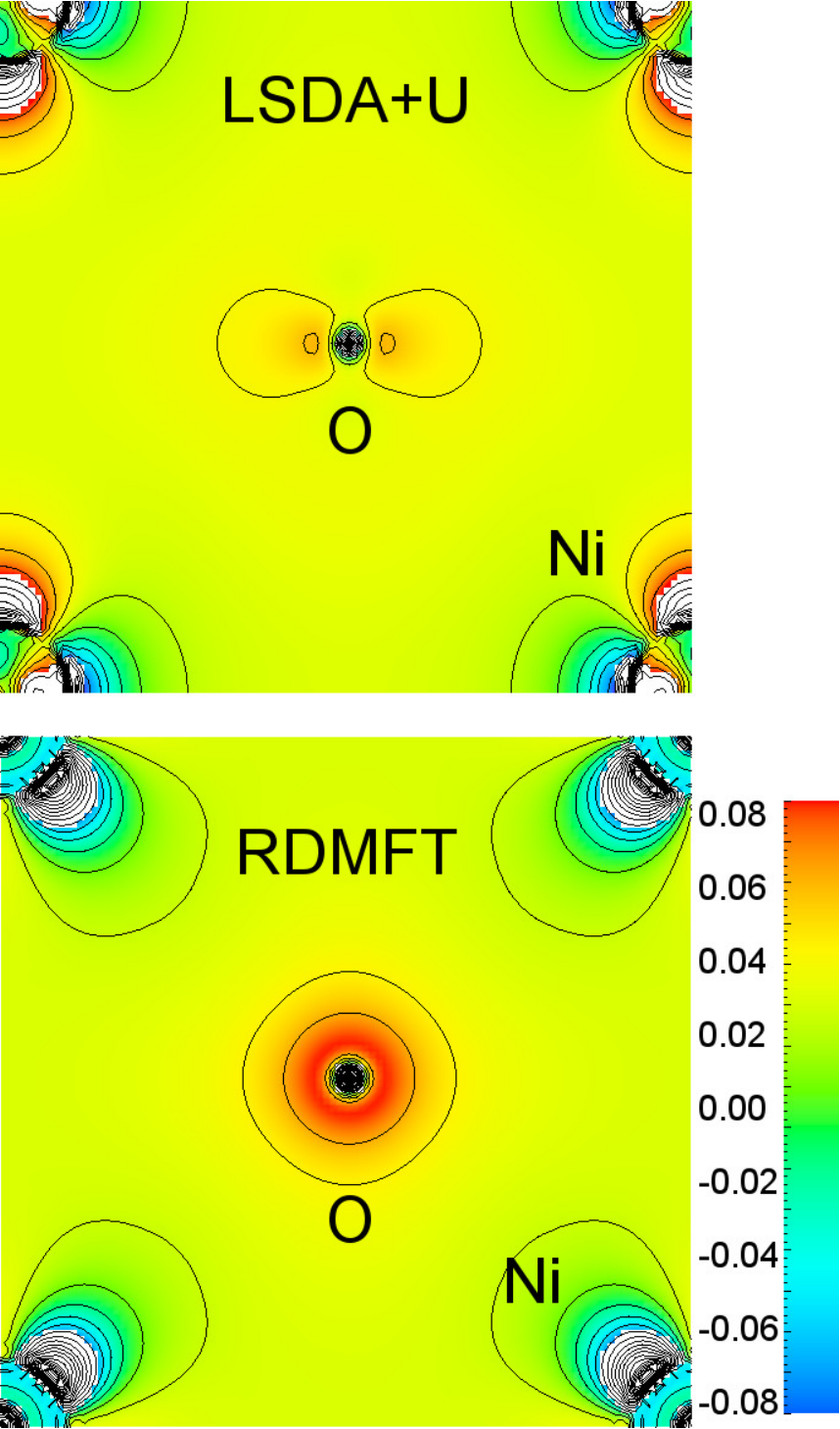}}
\caption{(Color online) Difference between the LSDA charge density and charge density 
calculated using LSDA+$U$ and RDMFT, ($\rho({\bf r})-\rho_{LSDA}({\bf r})$) for 
NiO. Positive values indicate localization of charge as compared to LSDA.}
\label{rho2d}
\end{figure}

A confirmation of this charge localization may be seen in the charge density.
In Fig. \ref{rho2d} we plot the difference $\rho({\bf r})-\rho_{LDA}({\bf r})$, 
for (i) RDMFT (lower panel) and (ii) the LSDA+$U$ functonal\cite{ldapu} (upper panel) within DFT for NiO. 
LSDA+$U$ method is chosen because, like RDMFT, it also finds the correct insulating ground state for NiO\cite{rodl09,nio-sharma}. 
The LSDA+$U$ method achieves this by both spin order an on-site Hubbard $U$ and, in contrast to RDMFT, 
cannot treat the non-magnetic insulating state of this material. The impact of this difference on the charge 
density is clear in Fig. \ref{rho2d}: significant charge localization is seen only in the RDMFT density. Interestingly,
one observes an almost spherical charge accumulation at the oxygen 
site, a result in agreement with experiment\cite{dudarev00}, but different
from that found in the corresponding LSDA+$U$ result.

%%%%%%%%%%%%%
% Conclusions
%%%%%%%%%%%%%
\section{Summary}

To summarize, in this work we demonstrate that RDMFT in conjunction with the power functional is able to capture the insulating state of
NiO and MnO in absence of long range spin order. However, while spin order does not drive the insulating ground state, 
the large local moments in these materials require spin
be explicitly taken into account for excellent agreement with experimental spectra to be obtained. The power, $\alpha$,
in the power-functional is an indicator of the amount of correlation and a detailed analysis 
shows a highly non trivial behaviour of the spectra, for Mott insulators, as a function of $\alpha$, which must lie 
within a small range (between 0.65 and 0.7) for the correct insulating ground-state is obtained. It is further 
shown that the natural orbitals for the strongly correlated materials, NiO and MnO, 
are much more localized as compared to the Kohn-Sham orbitals, which enables them to capture the physics of Mott localization in these materials.

%\bibliography{spectra}

\begin{thebibliography}{44}
\expandafter\ifx\csname natexlab\endcsname\relax\def\natexlab#1{#1}\fi
\expandafter\ifx\csname bibnamefont\endcsname\relax
  \def\bibnamefont#1{#1}\fi
\expandafter\ifx\csname bibfnamefont\endcsname\relax
  \def\bibfnamefont#1{#1}\fi
\expandafter\ifx\csname citenamefont\endcsname\relax
  \def\citenamefont#1{#1}\fi
\expandafter\ifx\csname url\endcsname\relax
  \def\url#1{\texttt{#1}}\fi
\expandafter\ifx\csname urlprefix\endcsname\relax\def\urlprefix{URL }\fi
\providecommand{\bibinfo}[2]{#2}
\providecommand{\eprint}[2][]{\url{#2}}

\bibitem[{\citenamefont{Kohn and Sham}(1965)}]{Kohn1965}
\bibinfo{author}{\bibfnamefont{W.}~\bibnamefont{Kohn}} \bibnamefont{and}
  \bibinfo{author}{\bibfnamefont{L.~J.} \bibnamefont{Sham}},
  \bibinfo{journal}{Phys. Rev.} \textbf{\bibinfo{volume}{140}},
  \bibinfo{pages}{A1133} (\bibinfo{year}{1965}).

\bibitem[{\citenamefont{Ulmer}(1959)}]{bis}
\bibinfo{author}{\bibfnamefont{K.}~\bibnamefont{Ulmer}},
  \bibinfo{journal}{Phys. Rev. Lett.} \textbf{\bibinfo{volume}{3}},
  \bibinfo{pages}{514} (\bibinfo{year}{1959}).

\bibitem[{\citenamefont{van Elp et~al.}(1991{\natexlab{a}})\citenamefont{van
  Elp, Potze, Eskes, Berger, and Sawatzky}}]{elp91a}
\bibinfo{author}{\bibfnamefont{J.}~\bibnamefont{van Elp}},
  \bibinfo{author}{\bibfnamefont{R.~H.} \bibnamefont{Potze}},
  \bibinfo{author}{\bibfnamefont{H.}~\bibnamefont{Eskes}},
  \bibinfo{author}{\bibfnamefont{R.}~\bibnamefont{Berger}}, \bibnamefont{and}
  \bibinfo{author}{\bibfnamefont{G.~A.} \bibnamefont{Sawatzky}},
  \bibinfo{journal}{Phys. Rev. B} \textbf{\bibinfo{volume}{44}},
  \bibinfo{pages}{1530} (\bibinfo{year}{1991}{\natexlab{a}}).

\bibitem[{\citenamefont{van Elp et~al.}(1991{\natexlab{b}})\citenamefont{van
  Elp, Wieland, Eskes, Kuiper, Sawatzky, de~Groot, and Turner}}]{elp91b}
\bibinfo{author}{\bibfnamefont{J.}~\bibnamefont{van Elp}},
  \bibinfo{author}{\bibfnamefont{J.~L.} \bibnamefont{Wieland}},
  \bibinfo{author}{\bibfnamefont{H.}~\bibnamefont{Eskes}},
  \bibinfo{author}{\bibfnamefont{P.}~\bibnamefont{Kuiper}},
  \bibinfo{author}{\bibfnamefont{G.~A.} \bibnamefont{Sawatzky}},
  \bibinfo{author}{\bibfnamefont{F.~M.~F.} \bibnamefont{de~Groot}},
  \bibnamefont{and} \bibinfo{author}{\bibfnamefont{T.~S.}
  \bibnamefont{Turner}}, \bibinfo{journal}{Phys. Rev. B}
  \textbf{\bibinfo{volume}{44}}, \bibinfo{pages}{6090}
  (\bibinfo{year}{1991}{\natexlab{b}}).

\bibitem[{\citenamefont{Sawatzky and Allen}(1984)}]{sawat84}
\bibinfo{author}{\bibfnamefont{G.~A.} \bibnamefont{Sawatzky}} \bibnamefont{and}
  \bibinfo{author}{\bibfnamefont{J.~W.} \bibnamefont{Allen}},
  \bibinfo{journal}{Phys. Rev. Lett.} \textbf{\bibinfo{volume}{53}},
  \bibinfo{pages}{2339} (\bibinfo{year}{1984}).

\bibitem[{\citenamefont{{Tjernberg {\it et al.}}}(1996)}]{tjer}
\bibinfo{author}{\bibfnamefont{O.}~\bibnamefont{{Tjernberg {\it et al.}}}},
  \bibinfo{journal}{Phys. Rev. B} \textbf{\bibinfo{volume}{54}},
  \bibinfo{pages}{10245} (\bibinfo{year}{1996}).

\bibitem[{\citenamefont{Jauch and Reehuis}(2004)}]{jauch}
\bibinfo{author}{\bibfnamefont{W.}~\bibnamefont{Jauch}} \bibnamefont{and}
  \bibinfo{author}{\bibfnamefont{M.}~\bibnamefont{Reehuis}},
  \bibinfo{journal}{Phys. Rev. B} \textbf{\bibinfo{volume}{70}},
  \bibinfo{pages}{195121} (\bibinfo{year}{2004}).

\bibitem[{\citenamefont{R\"odl et~al.}(2009)\citenamefont{R\"odl, Fuchs,
  Furthm\"uller, and Bechstedt}}]{rodl09}
\bibinfo{author}{\bibfnamefont{C.}~\bibnamefont{R\"odl}},
  \bibinfo{author}{\bibfnamefont{F.}~\bibnamefont{Fuchs}},
  \bibinfo{author}{\bibfnamefont{J.}~\bibnamefont{Furthm\"uller}},
  \bibnamefont{and}
  \bibinfo{author}{\bibfnamefont{F.}~\bibnamefont{Bechstedt}},
  \bibinfo{journal}{Phys. Rev. B} \textbf{\bibinfo{volume}{79}},
  \bibinfo{pages}{235114} (\bibinfo{year}{2009}).

\bibitem[{\citenamefont{Aryasetiawan and Gunnarsson}(1995)}]{arya95}
\bibinfo{author}{\bibfnamefont{F.}~\bibnamefont{Aryasetiawan}}
  \bibnamefont{and}
  \bibinfo{author}{\bibfnamefont{O.}~\bibnamefont{Gunnarsson}},
  \bibinfo{journal}{Phys. Rev. Lett.} \textbf{\bibinfo{volume}{74}},
  \bibinfo{pages}{3221} (\bibinfo{year}{1995}).

\bibitem[{\citenamefont{Kobayashi et~al.}(2008)\citenamefont{Kobayashi, Ohara,
  Yamamoto, and Fujiwara}}]{kobayashi08}
\bibinfo{author}{\bibfnamefont{S.}~\bibnamefont{Kobayashi}},
  \bibinfo{author}{\bibfnamefont{Y.}~\bibnamefont{Ohara}},
  \bibinfo{author}{\bibfnamefont{S.}~\bibnamefont{Yamamoto}}, \bibnamefont{and}
  \bibinfo{author}{\bibfnamefont{T.}~\bibnamefont{Fujiwara}},
  \bibinfo{journal}{Phys. Rev. B} \textbf{\bibinfo{volume}{78}},
  \bibinfo{pages}{155112} (\bibinfo{year}{2008}).

\bibitem[{\citenamefont{Ren et~al.}(2006)\citenamefont{Ren, Leonov, Keller,
  Kollar, Nekrasov, and Vollhardt}}]{ren06}
\bibinfo{author}{\bibfnamefont{X.}~\bibnamefont{Ren}},
  \bibinfo{author}{\bibfnamefont{I.}~\bibnamefont{Leonov}},
  \bibinfo{author}{\bibfnamefont{G.}~\bibnamefont{Keller}},
  \bibinfo{author}{\bibfnamefont{M.}~\bibnamefont{Kollar}},
  \bibinfo{author}{\bibfnamefont{I.}~\bibnamefont{Nekrasov}}, \bibnamefont{and}
  \bibinfo{author}{\bibfnamefont{D.}~\bibnamefont{Vollhardt}},
  \bibinfo{journal}{Phys. Rev. B} \textbf{\bibinfo{volume}{74}},
  \bibinfo{pages}{195114} (\bibinfo{year}{2006}).

\bibitem[{\citenamefont{Kunes et~al.}(2008)\citenamefont{Kunes, Lukoyanov,
  Anisimov, Scalettar, and Pickett}}]{kunes08}
\bibinfo{author}{\bibfnamefont{J.}~\bibnamefont{Kunes}},
  \bibinfo{author}{\bibfnamefont{A.~V.} \bibnamefont{Lukoyanov}},
  \bibinfo{author}{\bibfnamefont{V.~I.} \bibnamefont{Anisimov}},
  \bibinfo{author}{\bibfnamefont{R.~T.} \bibnamefont{Scalettar}},
  \bibnamefont{and} \bibinfo{author}{\bibfnamefont{W.~E.}
  \bibnamefont{Pickett}}, \bibinfo{journal}{Nat. Mat.}
  \textbf{\bibinfo{volume}{7}}, \bibinfo{pages}{198} (\bibinfo{year}{2008}).

\bibitem[{\citenamefont{Miura and Fujiwara}(2008)}]{oki08}
\bibinfo{author}{\bibfnamefont{O.}~\bibnamefont{Miura}} \bibnamefont{and}
  \bibinfo{author}{\bibfnamefont{T.}~\bibnamefont{Fujiwara}},
  \bibinfo{journal}{Phys. Rev. B} \textbf{\bibinfo{volume}{77}},
  \bibinfo{pages}{195124} (\bibinfo{year}{2008}).

\bibitem[{\citenamefont{Sharma et~al.}(2008)\citenamefont{Sharma, Dewhurst,
  Lathiotakis, and Gross}}]{sharma08}
\bibinfo{author}{\bibfnamefont{S.}~\bibnamefont{Sharma}},
  \bibinfo{author}{\bibfnamefont{J.~K.} \bibnamefont{Dewhurst}},
  \bibinfo{author}{\bibfnamefont{N.~N.} \bibnamefont{Lathiotakis}},
  \bibnamefont{and} \bibinfo{author}{\bibfnamefont{E.~K.~U.}
  \bibnamefont{Gross}}, \bibinfo{journal}{Phys. Rev. B}
  \textbf{\bibinfo{volume}{78}}, \bibinfo{pages}{201103}
  (\bibinfo{year}{2008}).

\bibitem[{\citenamefont{Sharma et~al.}(2013)\citenamefont{Sharma, Dewhurst,
  Shallcross, and Gross}}]{sharma13}
\bibinfo{author}{\bibfnamefont{S.}~\bibnamefont{Sharma}},
  \bibinfo{author}{\bibfnamefont{J.~K.} \bibnamefont{Dewhurst}},
  \bibinfo{author}{\bibfnamefont{S.}~\bibnamefont{Shallcross}},
  \bibnamefont{and} \bibinfo{author}{\bibfnamefont{E.~K.~U.}
  \bibnamefont{Gross}}, \bibinfo{journal}{Phys. Rev. Lett.}
  \textbf{\bibinfo{volume}{110}}, \bibinfo{pages}{116403}
  (\bibinfo{year}{2013}).

\bibitem[{\citenamefont{L\"odwin}(1955)}]{lodwin}
\bibinfo{author}{\bibfnamefont{P.~O.} \bibnamefont{L\"odwin}},
  \bibinfo{journal}{Phys. Rev.} \textbf{\bibinfo{volume}{97}},
  \bibinfo{pages}{1974} (\bibinfo{year}{1955}).

\bibitem[{\citenamefont{Gilbert}(1975)}]{gilbert}
\bibinfo{author}{\bibfnamefont{T.~L.} \bibnamefont{Gilbert}},
  \bibinfo{journal}{Phys. Rev. B} \textbf{\bibinfo{volume}{12}},
  \bibinfo{pages}{2111} (\bibinfo{year}{1975}).

\bibitem[{\citenamefont{Coleman}(1963)}]{coleman}
\bibinfo{author}{\bibfnamefont{A.}~\bibnamefont{Coleman}},
  \bibinfo{journal}{Rev. Mod. Phys.} \textbf{\bibinfo{volume}{35}},
  \bibinfo{pages}{668} (\bibinfo{year}{1963}).

\bibitem[{\citenamefont{M\"uller}(1984{\natexlab{a}})}]{M1984}
\bibinfo{author}{\bibfnamefont{A.~M.~K.} \bibnamefont{M\"uller}},
  \bibinfo{journal}{Phys. Rev. A} \textbf{\bibinfo{volume}{105}},
  \bibinfo{pages}{446} (\bibinfo{year}{1984}{\natexlab{a}}).

\bibitem[{\citenamefont{Gritsenko
  et~al.}(2005{\natexlab{a}})\citenamefont{Gritsenko, Pernal, and
  Baerends}}]{GPB2005}
\bibinfo{author}{\bibfnamefont{O.}~\bibnamefont{Gritsenko}},
  \bibinfo{author}{\bibfnamefont{K.}~\bibnamefont{Pernal}}, \bibnamefont{and}
  \bibinfo{author}{\bibfnamefont{E.~J.} \bibnamefont{Baerends}},
  \bibinfo{journal}{J. Chem. Phys.} \textbf{\bibinfo{volume}{122}},
  \bibinfo{pages}{204102} (\bibinfo{year}{2005}{\natexlab{a}}).

\bibitem[{\citenamefont{Rohr et~al.}(2010)\citenamefont{Rohr, Toulouse, and
  Pernal}}]{pernal2010}
\bibinfo{author}{\bibfnamefont{D.~R.} \bibnamefont{Rohr}},
  \bibinfo{author}{\bibfnamefont{J.}~\bibnamefont{Toulouse}}, \bibnamefont{and}
  \bibinfo{author}{\bibfnamefont{K.}~\bibnamefont{Pernal}},
  \bibinfo{journal}{Phys. Rev. A} \textbf{\bibinfo{volume}{82}},
  \bibinfo{pages}{052502} (\bibinfo{year}{2010}).

\bibitem[{\citenamefont{Rohr et~al.}(2008)\citenamefont{Rohr, Pernal,
  Gritsenko, and Baerends}}]{AC3}
\bibinfo{author}{\bibfnamefont{D.~R.} \bibnamefont{Rohr}},
  \bibinfo{author}{\bibfnamefont{K.}~\bibnamefont{Pernal}},
  \bibinfo{author}{\bibfnamefont{O.~V.} \bibnamefont{Gritsenko}},
  \bibnamefont{and} \bibinfo{author}{\bibfnamefont{E.~J.}
  \bibnamefont{Baerends}}, \bibinfo{journal}{J. Chem. Phys.}
  \textbf{\bibinfo{volume}{129}}, \bibinfo{pages}{164105}
  (\bibinfo{year}{2008}).

\bibitem[{\citenamefont{Marques and Lathiotakis}(2008)}]{pade}
\bibinfo{author}{\bibfnamefont{M.~A.~L.} \bibnamefont{Marques}}
  \bibnamefont{and} \bibinfo{author}{\bibfnamefont{N.~N.}
  \bibnamefont{Lathiotakis}}, \bibinfo{journal}{Phys. Rev. A}
  \textbf{\bibinfo{volume}{77}}, \bibinfo{pages}{032509}
  (\bibinfo{year}{2008}).

\bibitem[{\citenamefont{Piris}(2006)}]{pnof1}
\bibinfo{author}{\bibfnamefont{M.}~\bibnamefont{Piris}}, \bibinfo{journal}{Int.
  J. Quant. Chem} \textbf{\bibinfo{volume}{106}}, \bibinfo{pages}{1093}
  (\bibinfo{year}{2006}).

\bibitem[{\citenamefont{Giesbertz et~al.}(2010)\citenamefont{Giesbertz,
  Gritsenko, and Baerends}}]{PINOS1}
\bibinfo{author}{\bibfnamefont{K.~J.~H.} \bibnamefont{Giesbertz}},
  \bibinfo{author}{\bibfnamefont{O.~V.} \bibnamefont{Gritsenko}},
  \bibnamefont{and} \bibinfo{author}{\bibfnamefont{E.~J.}
  \bibnamefont{Baerends}}, \bibinfo{journal}{Phys. Rev. Lett.}
  \textbf{\bibinfo{volume}{105}}, \bibinfo{pages}{013002}
  (\bibinfo{year}{2010}).

\bibitem[{\citenamefont{Piris et~al.}(2011)\citenamefont{Piris, Lopez,
  Ruip\'erez, Matxain, and Ugalde}}]{PNOF5}
\bibinfo{author}{\bibfnamefont{M.}~\bibnamefont{Piris}},
  \bibinfo{author}{\bibfnamefont{X.}~\bibnamefont{Lopez}},
  \bibinfo{author}{\bibfnamefont{F.}~\bibnamefont{Ruip\'erez}},
  \bibinfo{author}{\bibfnamefont{J.~M.} \bibnamefont{Matxain}},
  \bibnamefont{and} \bibinfo{author}{\bibfnamefont{J.~M.}
  \bibnamefont{Ugalde}}, \bibinfo{journal}{J. Chem. Phys.}
  \textbf{\bibinfo{volume}{134}}, \bibinfo{pages}{164102}
  (\bibinfo{year}{2011}).

\bibitem[{\citenamefont{Pernal}(2014)}]{KP2014}
\bibinfo{author}{\bibfnamefont{K.}~\bibnamefont{Pernal}}, \bibinfo{journal}{J.
  Chem. Theory Comput.} \textbf{\bibinfo{volume}{10}}, \bibinfo{pages}{4332}
  (\bibinfo{year}{2014}).

\bibitem[{\citenamefont{Lathiotakis
  et~al.}(2014{\natexlab{a}})\citenamefont{Lathiotakis, Helbig, Rubio, and
  Gidopoulos}}]{localrdmft}
\bibinfo{author}{\bibfnamefont{N.~N.} \bibnamefont{Lathiotakis}},
  \bibinfo{author}{\bibfnamefont{N.}~\bibnamefont{Helbig}},
  \bibinfo{author}{\bibfnamefont{A.}~\bibnamefont{Rubio}}, \bibnamefont{and}
  \bibinfo{author}{\bibfnamefont{N.~I.} \bibnamefont{Gidopoulos}},
  \bibinfo{journal}{Phys. Rev. A} \textbf{\bibinfo{volume}{90}},
  \bibinfo{pages}{032511} (\bibinfo{year}{2014}{\natexlab{a}}).

\bibitem[{\citenamefont{Piris et~al.}(2013)\citenamefont{Piris, Matxain, and
  Lopez}}]{piris_jcp2013}
\bibinfo{author}{\bibfnamefont{M.}~\bibnamefont{Piris}},
  \bibinfo{author}{\bibfnamefont{J.~M.} \bibnamefont{Matxain}},
  \bibnamefont{and} \bibinfo{author}{\bibfnamefont{X.}~\bibnamefont{Lopez}},
  \bibinfo{journal}{J. Chem. Phys.} \textbf{\bibinfo{volume}{139}},
  \bibinfo{pages}{234109} (\bibinfo{year}{2013}).

\bibitem[{\citenamefont{Giesbertz et~al.}(2009)\citenamefont{Giesbertz, Pernal,
  Gritsenko, and Baerends}}]{GPGB2009}
\bibinfo{author}{\bibfnamefont{K.~J.~H.} \bibnamefont{Giesbertz}},
  \bibinfo{author}{\bibfnamefont{K.}~\bibnamefont{Pernal}},
  \bibinfo{author}{\bibfnamefont{O.~V.} \bibnamefont{Gritsenko}},
  \bibnamefont{and} \bibinfo{author}{\bibfnamefont{E.~J.}
  \bibnamefont{Baerends}}, \bibinfo{journal}{J. Chem. Phys.}
  \textbf{\bibinfo{volume}{130}}, \bibinfo{pages}{114104}
  (\bibinfo{year}{2009}).

\bibitem[{\citenamefont{Piris et~al.}(2012)\citenamefont{Piris, Matxain, Lopez,
  and Ugalde}}]{PMLU2012}
\bibinfo{author}{\bibfnamefont{M.}~\bibnamefont{Piris}},
  \bibinfo{author}{\bibfnamefont{J.~M.} \bibnamefont{Matxain}},
  \bibinfo{author}{\bibfnamefont{X.}~\bibnamefont{Lopez}}, \bibnamefont{and}
  \bibinfo{author}{\bibfnamefont{J.~M.} \bibnamefont{Ugalde}},
  \bibinfo{journal}{J. Chem. Phys.} \textbf{\bibinfo{volume}{136}},
  \bibinfo{pages}{174116} (\bibinfo{year}{2012}).

\bibitem[{\citenamefont{Chatterjee and Pernal}(2012)}]{CP2012}
\bibinfo{author}{\bibfnamefont{K.}~\bibnamefont{Chatterjee}} \bibnamefont{and}
  \bibinfo{author}{\bibfnamefont{K.}~\bibnamefont{Pernal}},
  \bibinfo{journal}{J. Chem. Phys.} \textbf{\bibinfo{volume}{137}},
  \bibinfo{pages}{204109} (\bibinfo{year}{2012}).

\bibitem[{\citenamefont{van Meer et~al.}(2013)\citenamefont{van Meer,
  Gritsenko, Giesbertz, and Baerends}}]{MGGB2013}
\bibinfo{author}{\bibfnamefont{R.}~\bibnamefont{van Meer}},
  \bibinfo{author}{\bibfnamefont{O.~V.} \bibnamefont{Gritsenko}},
  \bibinfo{author}{\bibfnamefont{K.~J.~H.} \bibnamefont{Giesbertz}},
  \bibnamefont{and} \bibinfo{author}{\bibfnamefont{E.~J.}
  \bibnamefont{Baerends}}, \bibinfo{journal}{J. Chem. Phys.}
  \textbf{\bibinfo{volume}{138}}, \bibinfo{pages}{094114}
  (\bibinfo{year}{2013}).

\bibitem[{\citenamefont{Lathiotakis
  et~al.}(2014{\natexlab{b}})\citenamefont{Lathiotakis, Helbig, Rubio, and
  Gidopoulos}}]{localrdmftappl}
\bibinfo{author}{\bibfnamefont{N.~N.} \bibnamefont{Lathiotakis}},
  \bibinfo{author}{\bibfnamefont{N.}~\bibnamefont{Helbig}},
  \bibinfo{author}{\bibfnamefont{A.}~\bibnamefont{Rubio}}, \bibnamefont{and}
  \bibinfo{author}{\bibfnamefont{N.~I.} \bibnamefont{Gidopoulos}},
  \bibinfo{journal}{J. Chem. Phys.} \textbf{\bibinfo{volume}{141}},
  \bibinfo{pages}{164120} (\bibinfo{year}{2014}{\natexlab{b}}).

\bibitem[{\citenamefont{M\"uller}(1984{\natexlab{b}})}]{mueller}
\bibinfo{author}{\bibfnamefont{A.~M.~K.} \bibnamefont{M\"uller}},
  \bibinfo{journal}{Phys. Lett.} \textbf{\bibinfo{volume}{105A}},
  \bibinfo{pages}{446} (\bibinfo{year}{1984}{\natexlab{b}}).

\bibitem[{\citenamefont{Cs\'anyi and Arias}(2000)}]{csanyi}
\bibinfo{author}{\bibfnamefont{G.}~\bibnamefont{Cs\'anyi}} \bibnamefont{and}
  \bibinfo{author}{\bibfnamefont{T.~A.} \bibnamefont{Arias}},
  \bibinfo{journal}{Phys. Rev. B} \textbf{\bibinfo{volume}{61}},
  \bibinfo{pages}{7348} (\bibinfo{year}{2000}).

\bibitem[{\citenamefont{Gritsenko
  et~al.}(2005{\natexlab{b}})\citenamefont{Gritsenko, Pernal, and
  Baerends}}]{gritsenko}
\bibinfo{author}{\bibfnamefont{O.}~\bibnamefont{Gritsenko}},
  \bibinfo{author}{\bibfnamefont{K.}~\bibnamefont{Pernal}}, \bibnamefont{and}
  \bibinfo{author}{\bibfnamefont{E.~J.} \bibnamefont{Baerends}},
  \bibinfo{journal}{J. Chem. Phys.} \textbf{\bibinfo{volume}{122}},
  \bibinfo{pages}{204102} (\bibinfo{year}{2005}{\natexlab{b}}).

\bibitem[{\citenamefont{Herbert and Harriman}(2003)}]{herbert}
\bibinfo{author}{\bibfnamefont{J.~M.} \bibnamefont{Herbert}} \bibnamefont{and}
  \bibinfo{author}{\bibfnamefont{J.~E.} \bibnamefont{Harriman}},
  \bibinfo{journal}{Chem. Phys. Lett.} \textbf{\bibinfo{volume}{382}},
  \bibinfo{pages}{142} (\bibinfo{year}{2003}).

\bibitem[{\citenamefont{Lathiotakis et~al.}(2007)\citenamefont{Lathiotakis,
  Helbig, and Gross}}]{nekjel}
\bibinfo{author}{\bibfnamefont{N.~N.} \bibnamefont{Lathiotakis}},
  \bibinfo{author}{\bibfnamefont{N.}~\bibnamefont{Helbig}}, \bibnamefont{and}
  \bibinfo{author}{\bibfnamefont{E.~K.~U.} \bibnamefont{Gross}},
  \bibinfo{journal}{Phys. Rev. B} \textbf{\bibinfo{volume}{75}},
  \bibinfo{pages}{195120} (\bibinfo{year}{2007}).

\bibitem[{elk(2004)}]{elk}
 (\bibinfo{year}{2004}), \urlprefix\url{http://elk.sourceforge.net}.

\bibitem[{\citenamefont{Perdew and Wang}(1992)}]{lda}
\bibinfo{author}{\bibfnamefont{J.~P.} \bibnamefont{Perdew}} \bibnamefont{and}
  \bibinfo{author}{\bibfnamefont{Y.}~\bibnamefont{Wang}},
  \bibinfo{journal}{Phys. Rev. B} \textbf{\bibinfo{volume}{45}},
  \bibinfo{pages}{13244} (\bibinfo{year}{1992}).

\bibitem[{\citenamefont{Liechtenstein et~al.}(1995)\citenamefont{Liechtenstein,
  Anisimov, and Zaanen}}]{ldapu}
\bibinfo{author}{\bibfnamefont{A.~I.} \bibnamefont{Liechtenstein}},
  \bibinfo{author}{\bibfnamefont{V.~I.} \bibnamefont{Anisimov}},
  \bibnamefont{and} \bibinfo{author}{\bibfnamefont{J.}~\bibnamefont{Zaanen}},
  \bibinfo{journal}{Phys. Rev. B} \textbf{\bibinfo{volume}{52}},
  \bibinfo{pages}{R5467} (\bibinfo{year}{1995}).

\bibitem[{\citenamefont{Shinohara et~al.}()\citenamefont{Shinohara, Sharma,
  Dewhurst, Shallcross, Lathiotakis, and Gross}}]{nio-sharma}
\bibinfo{author}{\bibfnamefont{Y.}~\bibnamefont{Shinohara}},
  \bibinfo{author}{\bibfnamefont{S.}~\bibnamefont{Sharma}},
  \bibinfo{author}{\bibfnamefont{J.~K.} \bibnamefont{Dewhurst}},
  \bibinfo{author}{\bibfnamefont{S.}~\bibnamefont{Shallcross}},
  \bibinfo{author}{\bibfnamefont{N.~N.} \bibnamefont{Lathiotakis}},
  \bibnamefont{and} \bibinfo{author}{\bibfnamefont{E.~K.~U.}
  \bibnamefont{Gross}}, \urlprefix\url{http://arxiv.org/abs/1206.1713}.

\bibitem[{\citenamefont{Dudarev et~al.}(2000)\citenamefont{Dudarev, Peng,
  Savrasov, and Zuo}}]{dudarev00}
\bibinfo{author}{\bibfnamefont{S.~L.} \bibnamefont{Dudarev}},
  \bibinfo{author}{\bibfnamefont{L.-M.} \bibnamefont{Peng}},
  \bibinfo{author}{\bibfnamefont{S.~Y.} \bibnamefont{Savrasov}},
  \bibnamefont{and} \bibinfo{author}{\bibfnamefont{J.-M.} \bibnamefont{Zuo}},
  \bibinfo{journal}{Phys. Rev. B} \textbf{\bibinfo{volume}{61}},
  \bibinfo{pages}{2506} (\bibinfo{year}{2000}).

\end{thebibliography}
\end{document}